\documentclass[journal]{IEEEtran}
\ifCLASSINFOpdf
\else
\fi
\usepackage{graphicx,subfigure}
\usepackage[T1]{fontenc}
\usepackage{algorithmic}
\usepackage{algorithm}
\usepackage{lipsum}
\usepackage{mathtools}
\usepackage[justification=centering]{caption}
\usepackage{cite}
 \usepackage{ragged2e}
\usepackage{multirow}
\usepackage{pbox}
\usepackage{array}
\usepackage{booktabs}

\usepackage{xcolor,colortbl}
\definecolor{gray}{rgb}{0.9, 0.9, 0.9}
\usepackage{amssymb}
\usepackage{textcomp}
\usepackage{float}
\usepackage{url}
\usepackage{hyperref}
\usepackage{amsmath}
\makeatletter
\renewcommand\paragraph{\@startsection{paragraph}{4}{\z@}%
            {-2.5ex\@plus -1ex \@minus -.25ex}%
            {1.25ex \@plus .25ex}%
            {\normalfont\normalsize\bfseries}}
\makeatother
\providecommand{\keywords}[1]{\textbf{\textit{Index terms---}} #1}
\usepackage[english]{babel}
\usepackage{babel,blindtext}
\usepackage{pbox}
\usepackage{float}
\usepackage{marvosym}
\usepackage{booktabs}

\begin{document}

\title{Real-time Collision Handling in Railway Network: An Agent-based Approach}


\author{Poulami~Dalapati,
	Abhijeet~Padhy,
	Bhawana~Mishra,
        Animesh~Dutta,
        and~Swapan~Bhattacharya
\thanks{P. Dalapati, A. Padhy, B. Mishra, and A. Dutta are with the Department
of Information Technology, National Institute of Technology Durgapur, WB 713209,
India, e-mail: dalapati89@gmail.com, abhijeet.padhy@gmail.com, bhawana20130790@gmail.com, animeshnit@gmail.com.}
\thanks{S. Bhattacharya is with the Department of CSE, Jadavpur University, Kolkata, India.
}}


\maketitle

\begin{abstract}
Advancement in intelligent transportation systems with complex operations requires autonomous planning and management
to avoid collisions in day-to-day traffic.
As failure and/or inadequacy in traffic safety system are life-critical, such collisions must be detected and resolved in an efficient way to manage continuously
rising traffic.
In this paper, we address different types of collision scenarios along with their early detection and resolution techniques in a complex railway system. In order to handle collisions
dynamically in distributed manner, a novel agent based
solution approach is proposed using the idea of \emph{max-sum} algorithm, where each agent (train agent, station agent, and junction agent)
communicates and cooperates with others to generate a good feasible solution that keeps the system in a safe state, i.e., collision free.
We implement the proposed mechanism in \emph{J}ava \emph{A}gent \emph{DE}velopment \emph{F}ramework (JADE). The results are evaluated with exhaustive experiments and compared
with different existing collision handling  methods to show the efficiency of our proposed approach.

\end{abstract}

\section{Introduction}
\label{intro}
\begin{figure}[!htbp]
\centering
 \includegraphics[scale=0.7]{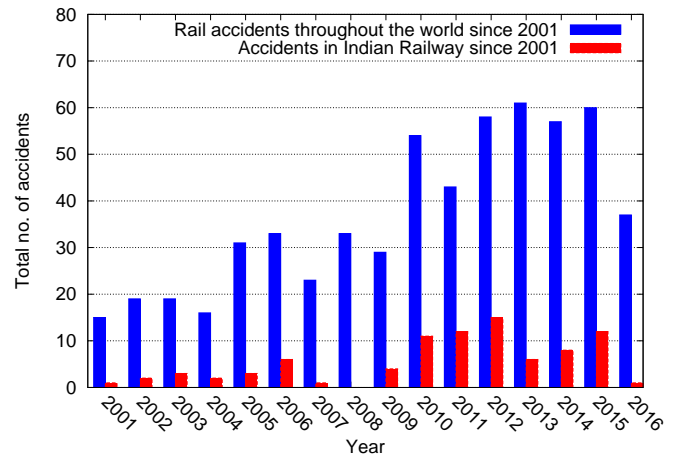}
 \captionsetup{justification=justified, singlelinecheck=false}
\caption{Overview of worldwide rail accidents since year 2001 \cite{collision_rail_statistics}.}
\label{fig:comparison}
\end{figure}

\begin{figure}[!htbp]
\centering
 \includegraphics[scale=0.7]{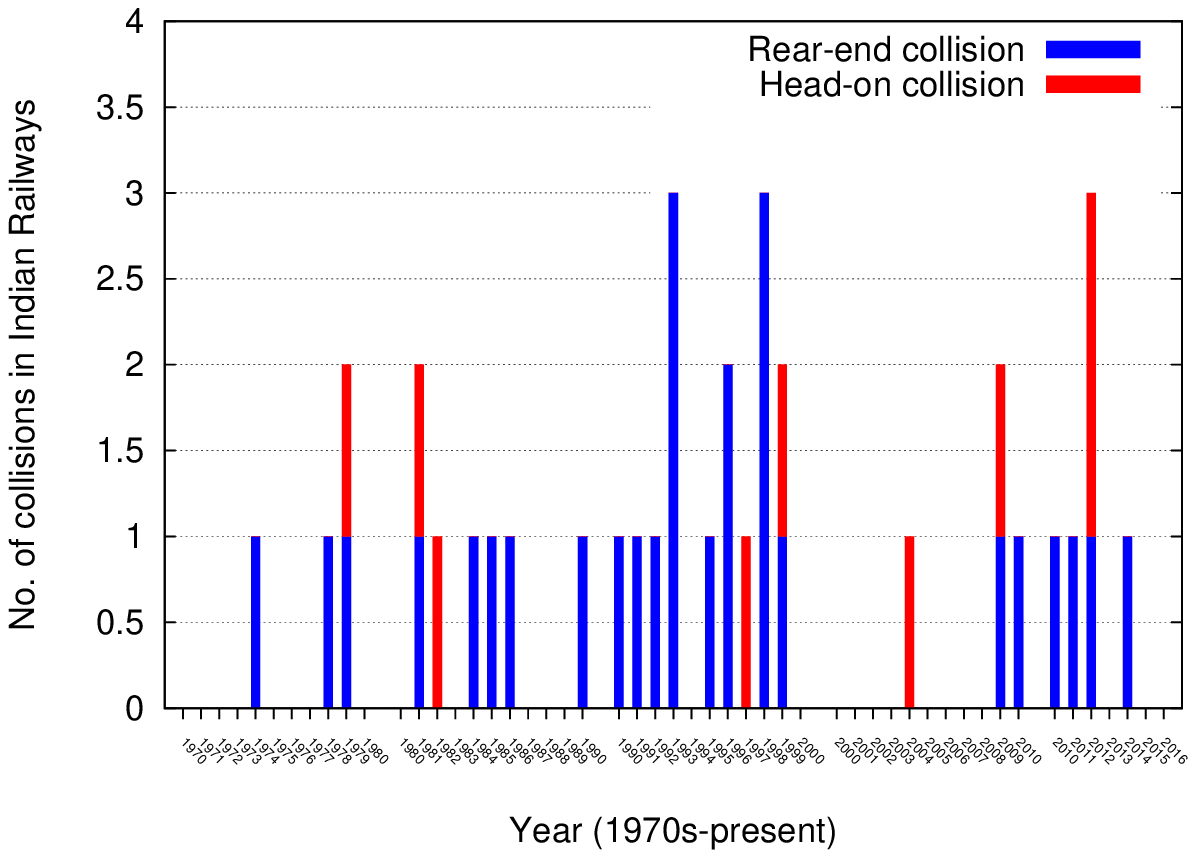}
 \captionsetup{justification=justified, singlelinecheck=false}
\caption{Overview of rear-end and head-on collisions in Indian Railway since 1970s \cite{rail_accidents_india}.}
\label{fig:collision_IR}
\end{figure}

Being the largest network in transportation systems \cite{prob425}, the railway system is very prone to collision \cite{collision, collision_india}.
In order to handle such situations, Advanced Train Control Systems (ATCS) \cite{ATCS} are being installed by railway authorities in various countries, which is mostly centralized.
Mainly to avoid collisions among trains, the railway system infrastructure consists of the comprehensive and complex technologies, such as train
control system with \emph{Automatic Block Signaling} (ABS) \cite{prob419} and \emph{interlocking} \cite{prob420, prob412}.
Despite the advancement in the technologies it is still found that, there exists an enormous number of collisions among trains
 \cite{collision_headon, collision_rearend} in different parts of the world.
 According to the data extracted from a large amount of historical data of accident statistics, shown in \emph{Figure \ref{fig:comparison}},
 it is noticed that, since the year 2001 significant number of accidents took place in railway, all over the world. Furthermore, every year
catastrophes like rear-end and head-on collisions are detected in Indian Railways (see \emph{Figure \ref{fig:collision_IR}}) with significant impact.
It is to be noted that, most of the rail accidents occur due to the human errors and the communication failure (or erroneous communication) between trains and control center.
 Additionally, only the operation center
has an overview of the rail traffic situation and based on the current traffic situation a train driver could only be intimated about the anticipated collision if an operation/control
 center can foresee it.

Earlier there had been some work in different transportation domains
\cite{prob401, prob402, prob403, prob404, prob405, prob406, prob407, prob409, prob410, prob411, prob412, prob415, prob421, prob422}
to achieve goals with different aspects using automated collision resolution techniques. In past, collision handling
in air traffic \cite{prob401} and road
traffic \cite{prob402, prob403, prob404, prob405, prob406, prob407, prob409} were under the microscopic lens of the researchers, whereas collision in railway transport
is not so widely nurtured till date \cite{prob410, prob411, prob412, prob415, prob421}. 
The present methods of controlling railway system are not able to handle the immense sensitivity arises due to an upsurge in day-to-day traffic and
complex operations to manage them.
Trains are manually controlled and operated by drivers, based on \emph{track-side interlocking} and \emph{blocking} with train signals and surveillance 
in conventional railway systems.
Moreover, the current scenario does not always allow direct train-to-train communication through message passing.
In case of crisis situations, trains need to contact its monitoring stations and final decisions taken by stations are conveyed to trains.
communication delay in severe case may increase the chances of collision.
Considering the above mentioned issues, it is necessary to develop a system that permits the trains to have an up-to-date and accurate information of the real-time
traffic situation in proximity, so that, the trains themselves can act accordingly to avoid dreadful accidents. Hence, our current work focuses on such collision handling in a complex
railway network.
For an immediate response of a collision scenario dynamically, the availability of timely and accurate information (position of train, speed of train, platform availability at stations,
 availability of junction at a time instant etc.) has a vital importance.
Moreover, to overcome the failure due to human error and hardship of centralized management, some level of autonomy in railway system is needed. 
Thus, the use of autonomous agents
(software agents with embedded sensor equipment) for communication and coordination in crisis situation (collision) has become a prime interest. The agents in multi-agent system 
can address entities in railway architecture like train, station, junction which can communicate among themselves to take a decision whenever needed. Here, the
proposed system is aimed not to rely on centralized infrastructure based control. It introduces each
train, station, and junction as an autonomous agent as the 
agents can much more accurately judge their positions, the distance among themselves, and the velocities of trains, can attentively monitor its surroundings and react instantly
to situations that would leave a human being helpless. 
the concept of agent-to-agent communication (train-to-train, train-to-station, train-to-junction) is introduced which
can ignore the need of track-side signaling.
Each agent communicates with nearby agents time to time and the neighborhood is determined by the communication range of individual agent.
With this concept, trains can take care of their safe distance and can generate alert at critical
situation. For this purpose, it is 
assumed that all the trains, stations, and junctions (cross-over point) are
equipped with communication devices of circular range.
The main challenges with such system model are that, the range of communication devices cannot cover entire region all at once.
Previous multi-agent based negotiation techniques are less sophisticated and less applicable for modeling
complex scenarios. 
Again, taking all these issues into consideration, mathematical modeling of such systems has increased concern for the safety of the system.
In this paper, our addressed collision scenario is broadly categorized into two main types: 
\emph{Head-on} collision, where front end of two trains collide and \emph{Rear-end} collision where a train smashes into the rear of other train. 
We propose collision detection and resolution techniques in railway system as a multi-agent based 
decentralized coordination. 
In collision detection phase, the system aims to detect the situation which may lead to a fatal collision. The goal of collision resolution is to prevent such destruction
to keep the system safe from the adversity. The idea of max-sum algorithm \cite{prob413, prob417} is used to generate the safe state, i.e.,
collision free system. The reason behind using 
the notion of max-sum is that, it can generate feasibly good solutions in such cases with less computation and communication.
Again, the manual calculation and computation in real-time for large complex network are very hard and time-consuming. So, to support the scalability dynamically,
an algorithmic approach is necessary.\\
\indent
In light of the discussion above, the \emph{main contributions} of this paper are,
\begin{itemize}
 \item In this paper, a multi-agent based model for collision handling problem is adopted which overcomes the need of track-side signaling systems and regular human 
 interventions.
 \item A \emph{max-sum} based decentralized solution approach with agent communication and negotiation is proposed to determine a safe state when any collision scenario
 is detected.
 \item Besides the modeling of the problem scenario and collision detection-resolution approach, the other aspects of this paper include the validation
 of proposed approaches
 and comparison with other existing approaches in similar domain.
\end{itemize}

The rest of the paper is organized as follows: In section \ref{state_of_art} some previous works
in related domain are summarized. Section \ref{system_model} is devoted to the description of railway network
and modeling of the system. 
Collision detection and resolution techniques are discussed in section \ref{collision_detection_resolution}.
Section \ref{results} highlights the experimental results and its validation in comparison with other existing approaches. 
Finally, section \ref{conclusion} concludes the proposed work with its future direction.\\

\section{State-of-the-Art}
\label{state_of_art}
In literatures \cite{prob401, prob402, prob403, prob404, prob405, prob406, prob407, prob409, prob410, prob411, prob412, prob414, prob415, prob416, prob421, wu2015modeling, wang2017novel, zhao2015positive, rajkumar2015approach, jain2012collision, dhanabalu2015sensor},
starting from air-traffic to road,
marine, and railway traffic, a variety of approaches for collision avoidance of vehicles in a complex environment have been proposed.
The recent improvement in such approaches has provided efficient algorithms that easily handle hundreds of vehicles, but cannot yet deal with
independent and autonomous agents of complex, realistic planning. Hence, it has been a major area of
interest for researchers from various fields and is still an active area of research.\\
\indent
In this section, some relevant approaches are discussed briefly.
D. Sislak et. al. in \cite{prob401} proposes two different implementation approaches
to the presented optimization-based collision avoidance in air traffic domain, parallel and semi-centralized, where airplanes
search for a series of actions that would allow them to avoid a
collision effectively. For the simplicity of description, conflict resolution actions have been limited to only horizontal control-
heading changes. However, the presented approach can be extended and actions can also include vertical and speed control,
if necessary. Here the proposed concept considers that all airplanes
can communicate and cooperate during conflict detection and avoidance phase. However,
the concept can be extended to include non-cooperative airplanes
flying in the same airspace.
Researches in \cite{prob402, prob403, prob405, prob406, prob407, prob409} highlight some motion models and collision avoidance approaches
to handle the possibility of unexpected maneuver. An idea of least restrictive supervisors for intersection
collision avoidance, for example, vehicle intersection crossing, is addressed in \cite{prob402}. Authors claim that this system guarantees
crossing safety (collision-free) and least restrictiveness (minimal intervening set). Choices of decisions are left on the vehicle-agents which cross
intersection while avoiding conflict. In their previous paper \cite{prob403} also, authors have dealt with multi-agent based collision avoidance. 
Some more research approaches in this similar domain are discussed in \cite{prob404, prob405}. Vehicle-to-vehicle (V2V) communication based technologies
\cite{prob406} and Automotive Collision Avoidance system \cite{prob407} are also nurtured to analyze worst case in collision avoidance systems.\\
\indent
In contrast to the presented approach, the above mentioned techniques are not suitable for railway systems. 
With railway, the main challenges lie into the spatial constraint as the rail wagon cannot divert horizontally even in unexpected situation. In case of emergency
situation (collision), a train driver can only brake or accelerate.
So, collision warning is one of the most important functions of railway safety systems.
Research papers \cite{prob410, prob411, prob412, prob415, prob421} highlight collision avoidance strategy in railway transport domain.
J. Lin et al. \cite{prob411} proposed an enhanced safety strategy for collision avoidance for train control system based on
direct V2V communication. Their system receives and evaluates the information broadcasted by other trains and then triggers collision
alerts when potential collision is detected. Isomorphic Markov Model is established in this regard. In contrast to their dynamic redundant
communication among trains, our presented approach considers minimum number of message passing, as the messages among trains are passed only
when trains are within the communication range, which may cause collision if no further step is taken. Since braking distance of trains
can be noticeably large, communication devices with sufficient range is taken.\\
\indent
Approaches for verification of safety properties along railway crossing region and decision taking in railway interlocking systems are addressed
in \cite{prob410, prob412}. In \cite{prob410}, the main objective is to model the control of railway crossing through a bottom up approach by
providing intelligence to trains so that collision along crossing is avoided. Authors in \cite{prob412} have proposed
a model to check safety within railway interlocking system in a large railway network. However, none of these two papers consider
 the collision scenarios along the railway tracks or at stations. So, in contrast, our presented approach handles both; 1) collision
 on tracks and 2) collision at stations in a global manner. Andreas Lehner et al. \cite{prob421} have presented a surveillance strategy concept
 for autonomous rail collision avoidance system, exploiting direct train-to-train communications. Their focus in this domain includes
 message broadcast rate in alert and advisory concept. Though they have investigated different scenarios in stations and shunting yards, 
 main line with high-speed services are not considered. In contrast, our method takes all these cases into account as potential threats may come in
 any part, all over the network.
 \\
 Some rear-end collision avoidance strategies are discussed in \cite{wu2015modeling, wang2017novel}. In \cite{wu2015modeling}, authors consider only one track
 in one-way to model and analyze rear-end collision. The collision avoidance parameters such as train distance control system, 
 train state communication-control system and danger alert system are assumed to be incorporated within the system. To avoid a rear-end collision due to erroneous commands from Automatic Train Protection (ATP) system,
 a parallel Centralized Traffic Control (CTC) and ATP based interval control is proposed in \cite{wang2017novel}. The idea of wild geese formation is 
 used here in which, the ATP controls the train interval as goose interval, adjusting the interval between two following trains locally; while
 CTC controls the same as goose line to keep the formation globally. CTC act as a centralized monitor in case of emergencies. Whereas in this work,
 the system is fully distributed where each train can communicate with other trains or stations or junctions
 to take dynamic decisions to avoid collision without any centralized system interventions. Moreover, both rear-end and head-on collision avoidance
 strategies are described in the present work. Here, not only a single track in single way, the whole network with both up and down direction in multiple
 tracks are taken into consideration.
 \\
\indent
 In \cite{zhao2015positive} authors have designed a system called Positive Train Control (PTC) to improve safety and efficiency of railway operations.
 They use advanced information technologies such as dynamic headway based on active communication (wireless communication and GPS) in order to properly
 monitor train separation or headways, avoid possible collision and improve safety. Though each train can choose its headway dynamically, but in case
 of emergency, centralized train control system intervene to give alerts to trains. Some more recent work are discussed in \cite{rajkumar2015approach, jain2012collision,
 dhanabalu2015sensor} for train collision avoidance.


\section{System Model and Problem Description}
\label{system_model}
\begin{figure}[!htbp]
\centering
\subfigure[]{\label{fig:railway}\includegraphics[scale=0.35]{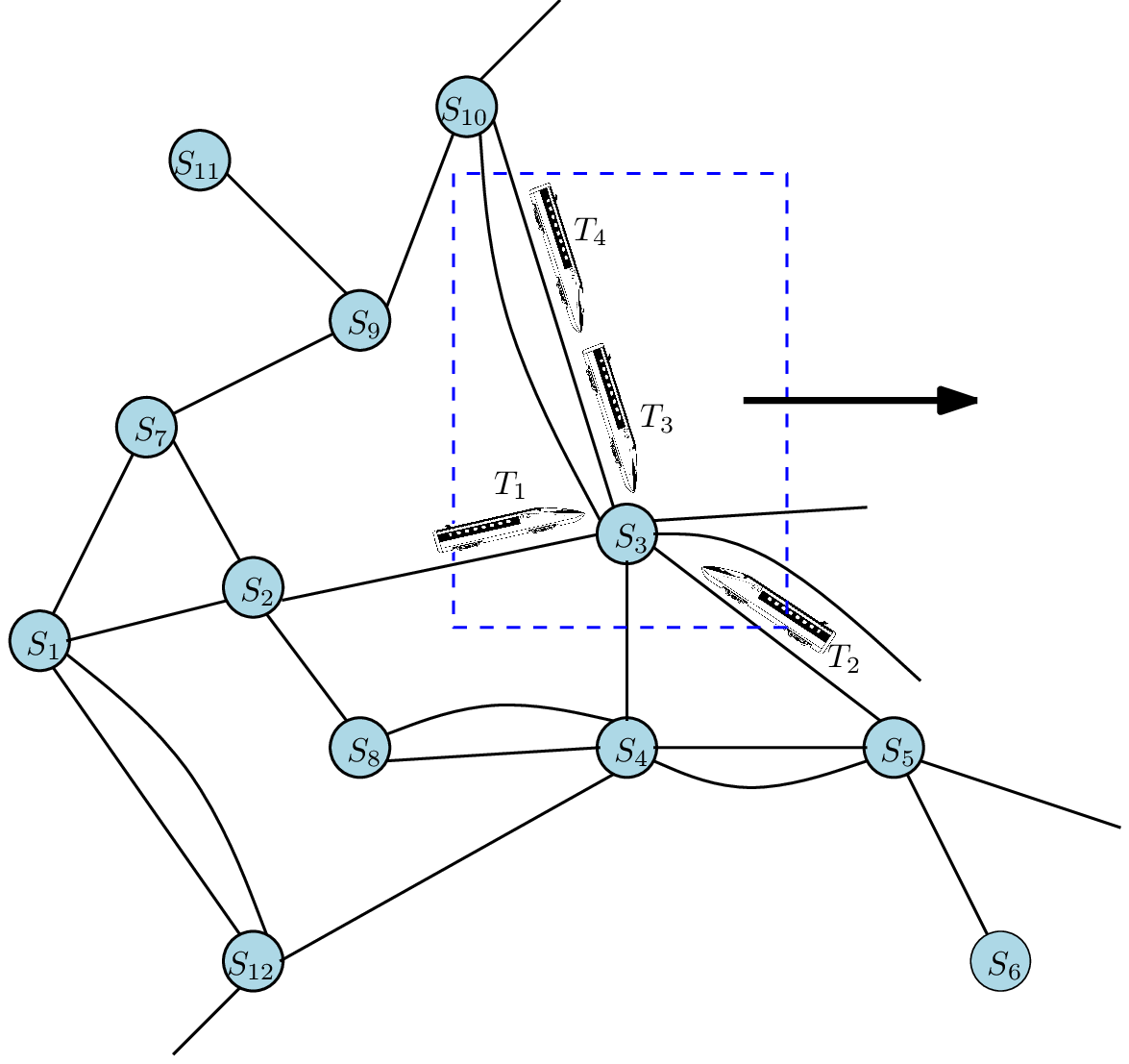}}
\subfigure[]{\label{fig:crossover}\includegraphics[scale=0.25]{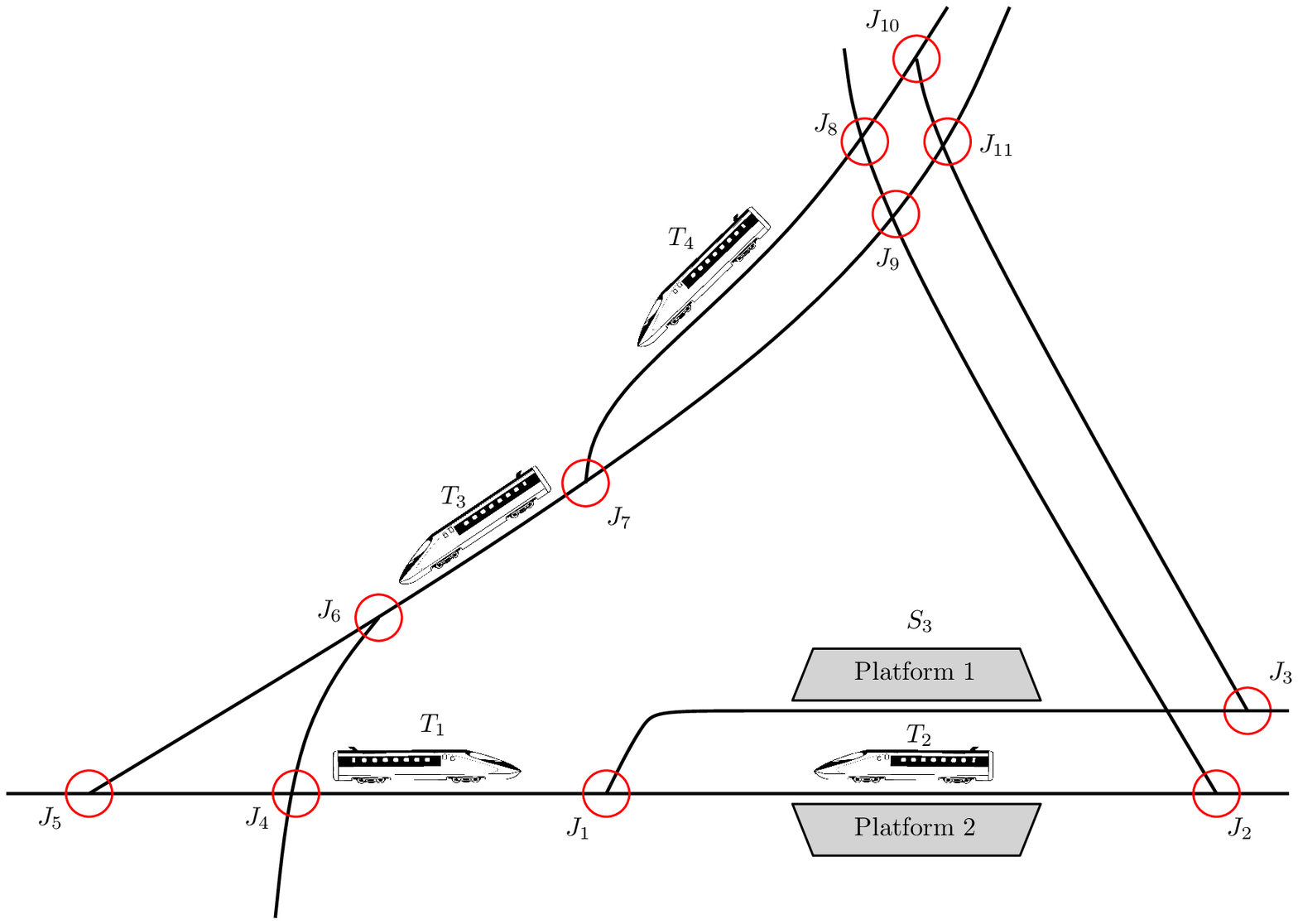}}
\captionsetup{justification=justified, singlelinecheck=false}
\caption{Railway Network. \newline
a Graphical representation of railway network.\newline
b Station with multiple tracks and in between cross-over (junction).}
\label{fig:RN}
\end{figure}

In the proposed model, we consider a railway network ($RN$) consisting of \emph{stations} ($S$), \emph{trains} ($T$),
\emph{junctions (cross-over)} ($\mathcal{J}$), and \emph{tracks}.
Multiple trains are either at stations or running on track at time instant $t$ as shown in \emph{Figure \ref{fig:RN}}.
Given this background, a \emph{Multi-agent System (MAS)} \cite{MAS, mas_fundamentals} is found to be suitable for 
modeling such distributed system.
Here, we represent $RN$ as a pair of multi-graph $\mathcal{G}$ and an agency $\mathcal{A}g$.
i.e., $RN = <\mathcal{G,A}g>$.
Again, $\mathcal{G=(V, E)}$, 
where $\mathcal{V}$ is set of vertices and $\mathcal{E}$ is set of edges and $\mathcal{A}g = \{ \mathcal{A}g_a | ~a\in [1, q] \}$, denotes agency.

\begin{table*}
 \caption{Notation}
\label{Notations}
\renewcommand{\arraystretch}{1.3}
\scalebox{0.8}{
\begin{tabular}{| l  p{5.5cm} || l  p{13cm} |}
\hline 
\rowcolor{gray}
\multicolumn{4}{|c|}{Indices and Parameters} \\
\hline \hline
$S$ & Stations & $b$ & Junction index, $b \subset i$ \\
$T$ & Trains & $j' \in [1,m]\backslash j$ & Index of train other than the $j^{th}$ train\\
$i$ & Station index & $i' \in [1,n]\backslash i$ & Index of station other than the $i^{th}$ station \\
$j$ & Train index & $a$ & Agent index\\
$l$ & Track index & $\rightarrow$ & Precedence relation between two trains, where $T_{j'}\rightarrow T_j$ implies $T_{j'}$ is following $T_j$\\
$\mathcal{J}$ & Junction & $t$ & Time instant\\
$x$ & Number of stations & $P_{jpl}$ & Platform indicator, $P_{jpl}=1$ if train $T_j$ is at $p^{th}$ platform occupying track $l$ \\
$m$ & Number of trains & $E|_t^{\mathcal{J}_b}$ & Total number of trains approaching towards junction $\mathcal{J}_b$ at time $t$, $1 \leq E|_t^{\mathcal{J}_b} \leq m$\\
$p$ & Platforms index & $d_j|_t^{\mathcal{J}_b}$ & Distance of train $T_j$ from junction $\mathcal{J}_b$ at time $t$\\
$d_{jj'}^{\mathcal{H}}$ & Headway between two train $T_j$ and $T_{j'}$ &$\vartheta_j|_t^l$ & Speed of train $T_j$ at time $t$ on track $l$ \\
$d_{j}^{\mathcal{B}}$ & Braking distance of train $T_j$ & $\partial_{T_j}|_{t}^{l}$ & Direction of train $T_j$ at time $t$ on track $l$,  $\partial_{T_j}|_{t}^{l}=1$ if $T_j$ runs in "UP" direction and 0 for "DOWN" direction\\
$d_{jj'}^{\mathcal{C}}$ & Critical distance between two train $T_j$ and $T_{j'}$ & $L_{jl}$ & Track indicator, $L_{jl}=1$ if train $j$ occupies $l^{th}$ track, otherwise 0 and when $L_{jl}=1$, $L_{j'l}=0$\\
$r$ & Range of the communication device & $c$ & Number of junctions\\
\hline
\end{tabular}
}
\end{table*}

\indent From notations in Table \ref{Notations},
$\mathcal{V} = \{v_g | ~g \in [1, n]\}$. 
Here $v_g$ can either be an element of station set $S = \{S_i | ~i \in [1,x]\}$ or an element of junction set $\mathcal{J} = \{\mathcal{J}_b | ~b \in [1,c]\}$.
$v_i = S_i$ means vertex is a station, $v_i= \mathcal{J}_i$ means vertex is a junction point
Again, $(S \cap \mathcal{J})=\phi$ and $x+c=n$.
$\mathcal{E}$ represents tracks between two stations or between two junctions or between station and junction. 
$T = \{T_j | ~j \in [1, m]\}$, indicates trains.
For example, in \emph{Figure \ref{fig:railway}} we have considered a railway network with $12$ stations and $4$ trains, where some trains are 
running on tracks ($T_1, T_3, T_4$) and some are standing at stations ($T_2$). With such scenario, the station $S_3$ and its surroundings are magnified in 
\emph{Figure \ref{fig:crossover}}. Here, train $T_2$ is standing at platform $2$ of station $S_3$, train $T_1$ has left previous station $S_2$
and is coming at the same platform at the same time, leading to the chances of collision. Again, train $T_3$ and $T_4$ have left station
$S_{10}$ and approaching to station $S_3$. Here, train $T_4$ is following train $T_3$.
The position of junctions $\mathcal{J}_1, \mathcal{J}_2, \ldots , \mathcal{J}_{10}$ are shown in \emph{Figure \ref{fig:crossover}}. Potential collision
may happen between $T_3$ and $T_4$ if there arise some speed discrimination between them.
In this paper each of the stations $S$, junctions $\mathcal{J}$ and trains $T$
are associated with agents named station agent $(SA)$, junction agent $(JA)$ and train agent $(TA)$ respectively such that the total number of agents $q=n+m$.
In this system model, $SA$, $JA$, and $TA$ all communicate and 
cooperate with each other to take decisions, ensuring distributed control and autonomy, to keep the railway system in a safe state, i.e. collision free.
Here, the basic idea of
\emph{headway distance} \cite{prob418}, \emph{braking distance} \cite{braking_distance}, and \emph{critical distance} \cite{critical_distance}
are taken as metric parameter to detect alarming situation in collision
scenario and stopping condition for trains to avoid collision.

\begin{figure}[!htbp]
\centering
 \includegraphics[scale=0.55]{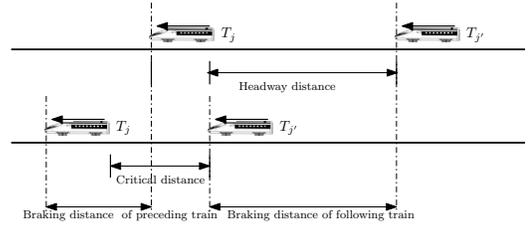}
 \captionsetup{justification=justified, singlelinecheck=false}
\caption{Headway distance, braking distance, and critical distance in rail transit.}
\label{fig:various_distance}
\end{figure}

\emph{Headway distance ($d_{jj'}^{\mathcal{H}}$)} is the minimum possible distance between two trains in transit.
It is measured as the distance from the tail of the preceding train to front of the following train (\emph{tip-to-tail measurements})
as depicted in \emph{Figure \ref{fig:various_distance}} for trains $T_j$ and $T_{j'}$.\\

\emph{Braking distance ($d_{j}^{\mathcal{B}}$)} refers to the distance traveled by the train from the point when its brakes are fully applied 
till it comes to a complete stop.
From the example described in \emph{Figure \ref{fig:various_distance}} let us assume that, both the trains $T_j$ and $T_{j'}$ apply full brake at time instant $t$.
The distance covered by them till it reaches to static state depends on the initial speed of train when the brake is applied,
and the coefficient of friction between the train wheels and the railway track. For any train $T_j$ it is calculated as:\\
\begin{equation*}
\label{eq:braking_distance}
 d_{j}^{\mathcal{B}}~=~\frac{(\vartheta_{j})^{2}}{2\mu_k g}
\end{equation*}
where,  $\vartheta_j$ = speed of train $T_j$, $g$ = gravity of earth, and $\mu_k$ = coefficient of kinetic friction.
As both the railway tracks and rail wheels are made up of high quality steel, the values of $\mu_k$ is taken as $\mu_k$ = 0.42.\\

\begin{figure}[!htbp]
\centering
\subfigure[]{\label{fig:Headway1}\includegraphics[scale=0.5]{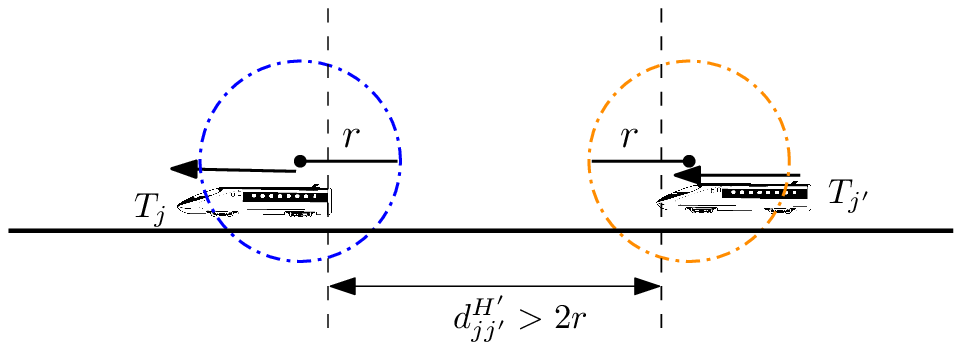}}
\subfigure[]{\label{fig:Headway2}\includegraphics[scale=0.5]{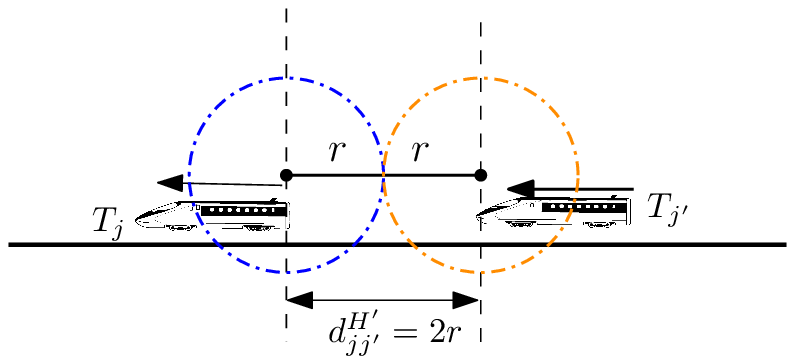}}
\hspace*{0.5cm}
\subfigure[]{\label{fig:Headway3}\includegraphics[scale=0.5]{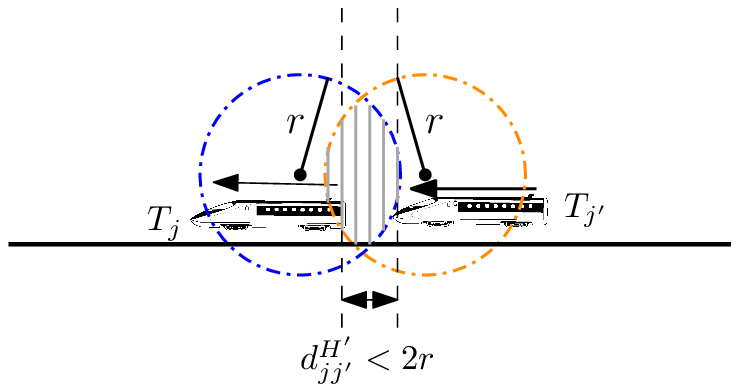}}
\captionsetup{justification=justified, singlelinecheck=false}
\caption{Relation between actual distance and communication range in collision scenarios. \newline
a Headway distance is much greater than two trains' communication range.\newline 
b Headway distance is equal to two trains' communication range.\newline 
c Headway distance is less than two trains' communication range.}
 \label{fig:Headway}
\end{figure}

\emph{Critical distance ($d_{jj'}^{\mathcal{C}}$)} is defined as the maximum acceptable gap between two trains, beyond which collision is inevitable.
It is also measured following \emph{tip-to-tail measurement} concept.
In general, critical distance of two trains is always less than the headway distance between them as depicted in \emph{Figure \ref{fig:various_distance}}.
\\
\indent
For direct communication between two trains they must be within their communication range, i.e. the maximum distance between them ($d_{jj'}^{\mathcal{H'}}$)
must be less than or equal to $2r$. If their current distance is greater than $2r$ then two trains cannot communicate directly using their communication
devices. There can be three such situations such as, current distance between two trains is greater than $2r$, 
current distance between two trains is equal to $2r$, current distance between two trains is less than $2r$ as described in \emph{Figure \ref{fig:Headway}}.

\section{Collision Detection and Resolution}
\label{collision_detection_resolution}
\subsection{Collision detection in a real-time scenario}
\label{detection}
In this paper, we consider the collision scenario with more than one trains as shown in \emph{Figure \ref{fig:headon}} and \emph{Figure \ref{fig:rearend}}.
It may be the case that, \emph{(i)} two trains are running on the same track,
but in opposite direction (see \emph{Figure \ref{fig:H2}}), \emph{(ii)} two trains are approaching to the same junction (cross-over point) at the same time
(see \emph{Figure \ref{fig:H1}}),
\emph{(iii)} two trains are running on the same track in same direction, but the speed of the following train is greater than the previous train and their headway distance
decreases to critical distance (see \emph{Figure \ref{fig:R2}}), \emph{(iv)} one train is standing at a platform of a station and another train is coming to the same platform at the same time
(see \emph{Figure \ref{fig:R1}}). Depending upon the situations discussed above,
railway collision is classified here into two types: \emph{(a)} Head-on collision and \emph{(b)} Rear-end collision.

\begin{itemize}
\item Case 1.1: \emph{Head-on Collision}\\
 A head-on collision is defined as a collision where front end of two trains hit each other due to some erroneous instructions outputted by the controlling authority.
 In railway system this kind of catastrophe arises in two situations: i) when two trains are moving forward on the same track in opposite direction,
 ii) both the trains are trying to cross the same junction at the same time, following some erroneous signal information.
 In such case, the distance required for a train to stop is usually greater than their sighting distance (i.e., when two trains become visible to
 each other), which leads to a fatal collision. Both the above mentioned cases are depicted in Figure
 \ref{fig:headon}, where in \emph{Figure \ref{fig:H2}} two trains $T_1$ and $T_2$ are running on same track at the same time with opposite direction.
 In \emph{Figure \ref{fig:H1}}, both
 $T_1$ and $T_2$ are approaching towards same junction $\mathcal{J}_b$ at same time. 
 The generalization of detection of head-on collision scenario is formulated as follows.
\begin{figure}[!htbp]
\centering
\subfigure[]{\label{fig:H2}\includegraphics[scale=0.45]{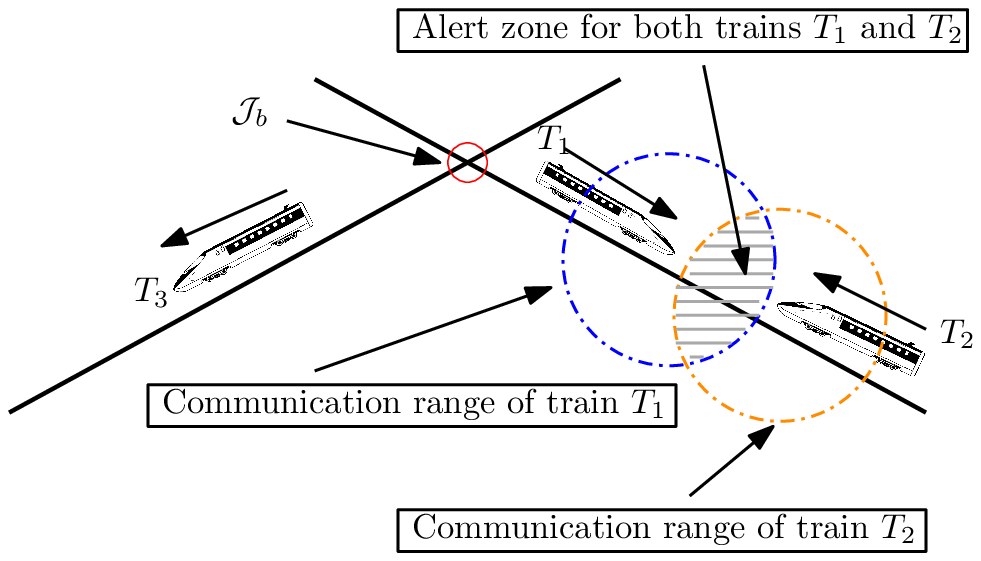}}
\subfigure[]{\label{fig:H1}\includegraphics[scale=0.45]{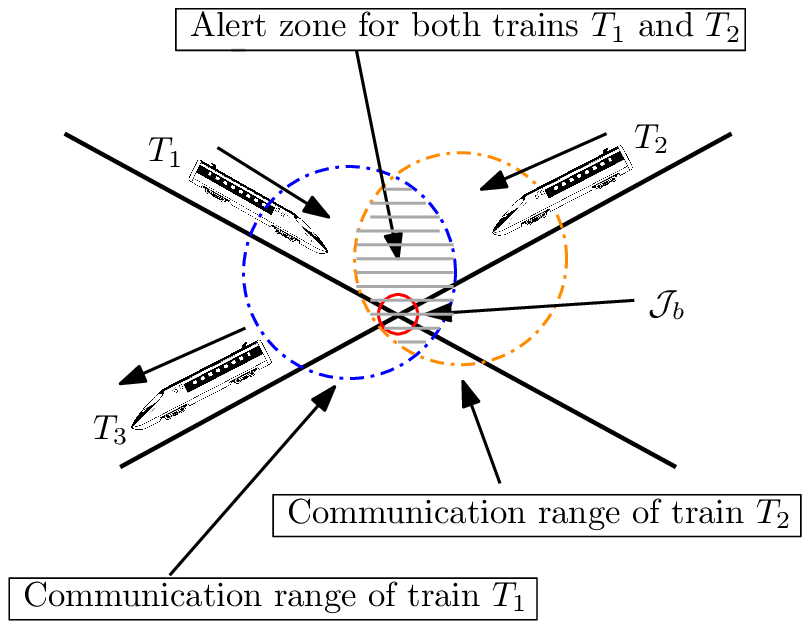}}
\captionsetup{justification=justified, singlelinecheck=false}
\caption{Head-on collision. \newline
a Two trains are running on the same track at same time, but with different directions.\newline
b Two trains are approaching to the same crossover point at the same time.}
\label{fig:headon}
\end{figure}

 \begin{itemize}
  \item Case 1.1.1: 
 Two trains $T_j$ and $T_{j'}$ are running on the same track $l$. i.e.,
  \begin{equation}
  \label{eq:1.1.1:1}
  \begin{cases}
    L_{jl} = 1 \\
    L_{j'l} = 1
  \end{cases}
  \end{equation}
 Both $T_j$ and $T_{j'}$ are in opposite direction on same track $l$ at same time $t$. i.e.,
  \begin{equation}
  \label{eq:1.1.1:2}
  \partial_{T_j}|_{t}^{l} \oplus \partial_{T_{j'}}|_{t}^{l} = 1 
  \end{equation}
  At time instant $t$, either of the trains or both the trains $T_j$ and $T_{j'}$ are in running state on track $l$, i.e.
  \begin{equation}
  \label{eq:1.1.1:3}
  \begin{cases}
  (\vartheta_j|_t^l > 0) ~and~ (\vartheta_{j'}|_t^l  = 0) \\
  or \\
  (\vartheta_j|_t^l = 0) ~and~ (\vartheta_{j'}|_t^l  > 0) \\
  or \\
  (\vartheta_j|_t^l > 0) ~and~ (\vartheta_{j'}|_t^l  > 0) \\
  \end{cases}
  \end{equation}
  and the current distance between the two trains $T_j$ and $T_{j'}$ at time instant $t$ is less than their predefined headway distance. i.e.,
  \begin{equation}
  \label{eq:1.1.1:4}
   d_{jj'}^{\mathcal{H'}}<d_{jj'}^{\mathcal{H}}
  \end{equation}
   
  \item Case 1.1.2: 
 Two trains $T_j$ and $T_{j'}$ are on different track $l$ and $l'$ respectively and approaching
  towards a same junction $\mathcal{J}_b$ at the same time $t$, i.e.,
  \begin{equation}
  \label{eq:1.1.2:1}
  \begin{cases}
   L_{jl}=1 \\
   L_{j'l'}=1
  \end{cases}
  \end{equation}
  and 
  \begin{equation}
  \label{eq:1.1.2:2}
  \begin{cases}
   \vartheta_j|_t^l > 0 \\
  \vartheta_{j'}|_t^{l'}  > 0
  \end{cases} 
  \end{equation}
  Junction $\mathcal{J}_b$ detects more than one entry of trains within its communication range $r'$, i.e.
  \begin{align}
  \label{eq:1.1.2:3}
   d_j|_t^{\mathcal{J}_b} \leq r \nonumber \\
   d_{j'}|_t^{\mathcal{J}_b} \leq r \\
   1 < E|_t^{\mathcal{J}_b} \leq m \nonumber
  \end{align}
  \end{itemize}
It is to be noted that, anyone of the above discussed cases (case 1.1.1 and case 1.1.2) may lead to a collision scenario in railway system.
 In their course of journey both the trains $T_j$ and $T_{j'}$ as well as junction $\mathcal{J}_b$ check for more than one entry of trains within their communication
 range. If current distance between two trains is greater than the sum of their individual range, i.e. $d_{jj'}^{\mathcal{H'}} > 2r$, then trains cannot communicate
  with each other directly to take cooperative decisions.
  Hence, the monitoring station $S_i$ communicates with both trains $T_j$ and $T_{j'}$ to take decisions as soon as possible to achieve a safe state.
  If distance between $T_j$ and $T_{j'}$ are within the communication range of each other, i.e. $d_{jj'}^{\mathcal{H'}} \leq 2r$,
  then an alert situation is detected and both the trains communicate directly to avoid collision. For \emph{case 1.1.1}, both
  $T_j$ and $T_{j'}$ apply full brake. If both of them can make a stop at a distance
  greater than or equal to the critical distance then the collision is avoided. 
  \begin{equation}
   \label{eq:5}
   (d_{jj'}^{\mathcal{H'}} - (d_{j}^{\mathcal{B}} + d_{j'}^{\mathcal{B}})) \ge  d_{jj'}^{\mathcal{C'}}
  \end{equation}
  So, to avoid a collision equation (\ref{eq:5}) must be satisfied. Otherwise a head-on collision is inevitable.
  For \emph{case 1.1.2}, primarily both $T_j$ and $T_{j'}$
  calculate instantly their braking distance $d_{j}^{\mathcal{B}}$ and $d_{j'}^{\mathcal{B}}$ respectively. Let us assume that, the time taken to calculate
  this is $\Delta t$ which is very small (tends to $0$) and the distance $\Delta d_j$ and $\Delta d_{j'}$ covered by trains $T_j$ and $T_{j'}$  respectively 
  within $\Delta t$ period of time is also negligibly small and do not hamper the braking distance; i.e.\\
  \begin{align}
  \label{eq:1.1:1}
  \begin{cases}
  d_{j}^{\mathcal{B}}+\Delta d_j \approx d_{j}^{\mathcal{B}} \\
  d_{j'}^{\mathcal{B}}+\Delta d_{j'} \approx d_{j'}^{\mathcal{B}}\\
  t+\Delta t \approx t
  \end{cases}
  \end{align} 
  where,
  \begin{align}
  \label{eq:1.1:1}
  \begin{cases}
  \Delta d_j=\vartheta_j \times \Delta t\\
 \Delta d_{j'}=\vartheta_{j'} \times \Delta t 
  \end{cases}
  \end{align} 
  Now presume that two trains $T_j$ and $T_{j'}$ are approaching to the same junction $\mathcal{J}_b$ at same time.
  If the distance of both $T_j$ and $T_{j'}$ from the junction $\mathcal{J}_b$ at time instant $t$ is greater than their respective braking distance, i.e.
  $d_j|_t^{\mathcal{J}_b}>d_{j}^{\mathcal{B}}$ and $d_{j'}|_t^{\mathcal{J}_b}>d_{j'}^{\mathcal{B}}$, then the high priority train (let say $T_j$) proceeds towards
  $\mathcal{J}_b$ and the other one (here $T_{j'}$) applies brake to stop. Otherwise, if braking distance of one 
  of the two trains is greater than its distance from the junction,
  then that train (let say $T_{j'}$) decides to use the junction first and the other train $T_j$ applies full brake to stop. But in
  worst case, if both $T_j$ and $T_{j'}$ have braking
  distance greater than their distance from crossover point then collision occurs. The priority is given to trains depending upon their category (Premium
  trains, Superfast Express, Express and Mail trains, Passenger and Fast Passenger, Freight trains etc., 
  where Premium trains hold highest priority and so on \cite{train_priority}).
 
  \item Case 1.2: \emph{Rear-end Collision}\\
  A rear-end collision in railway is defined as an accident where a train crashes into the rear of its preceding train. Typical scenarios for rear-end collisions are
  a sudden deceleration by the first train, so that the following train does not have enough braking distance and collides with the first.
  Alternatively, the following train may accelerate more rapidly than the preceding one, resulting a collision. Again it may be the case that, due to
  signaling error or human error or communication failure, a train comes to the same platform when another train is already standing there.\\
  \indent 
  All these cases are pictorially described in \emph{Figure \ref{fig:rearend}}, where in \emph{Figure \ref{fig:R2}}, two trains $T_1$ and $T_2$ are running on the same track
  in same direction and $T_2$ is following $T_1$. In such scenario, if either $T_1$ decelerate or $T_2$ accelerate or both happens together then the system detects collision.
  \emph{Figure \ref{fig:R1}} depicts the case where $T_1$ is already at station $S_1$ and $T_2$ is coming at the same platform of station $S_1$ at same time. These cases are mathematically 
  formulated below.
\begin{figure}[!htbp]
\centering
\subfigure[]{\label{fig:R2}\includegraphics[scale=0.45]{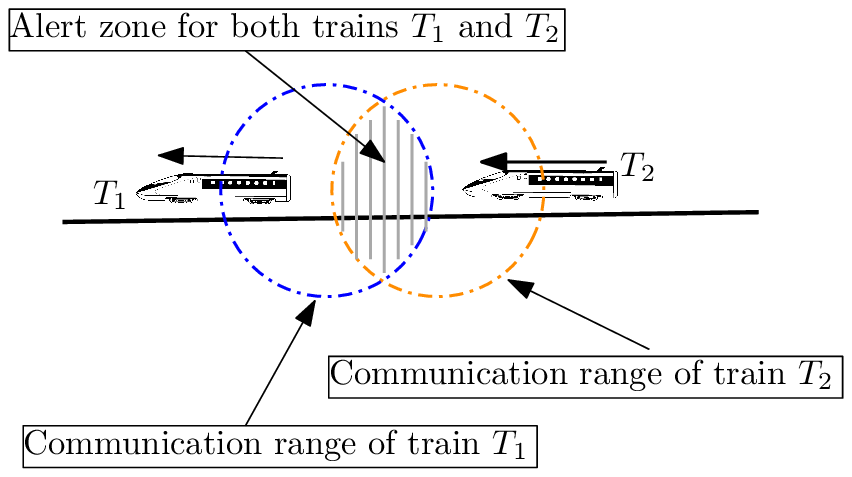}}
\subfigure[]{\label{fig:R1}\includegraphics[scale=0.45]{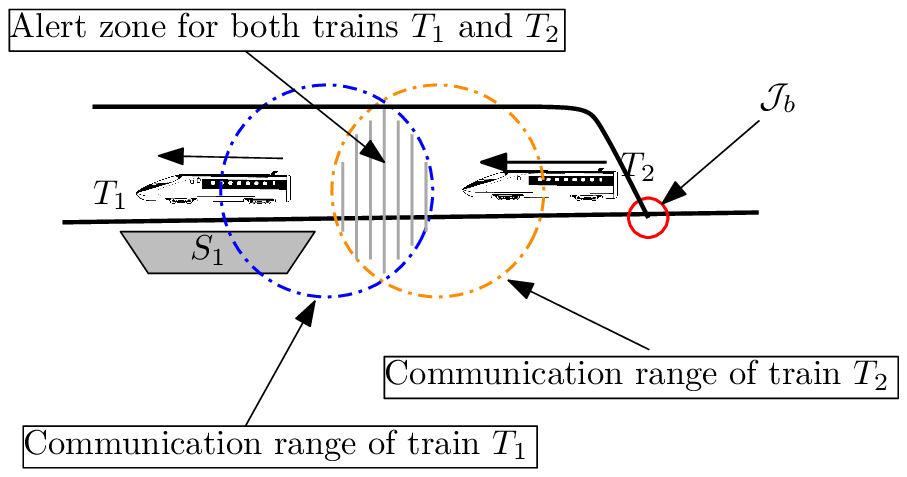}}
\captionsetup{justification=justified, singlelinecheck=false}
\caption{Rear-end collision. \newline
a Two trains are running on the same track at the same time, in the same directions.\newline
b One train is standing at platform and the following train is moving towards it on the same track.}
\label{fig:rearend}
\end{figure}

 \begin{itemize}
  \item Case 1.2.1: Two trains $T_j$ and $T_{j'}$ are running on the same track $l$. i.e.,   
  \begin{equation}
   \label{eq:1.2.1:1}
   \begin{cases}
   L_{jl} \\
   L_{j'l} = 1 
   \end{cases}
  \end{equation}
  Train $T_{j'}$ is following the train $T_j$ and the speed of the train $T_{j'}$ is greater than the preceding train $T_j$. i.e.,
  \begin{equation}
  \label{eq:1.2.1:2}
  \begin{cases}
  T_{j'}\rightarrow T_j  \\
  \partial_{T_j}|_{t}^{l} \odot \partial_{T_{j'}}|_{t}^{l} = 1 \\
  \vartheta_{j'} > \vartheta_j > 0
  \end{cases}
  \end{equation}
 and the current distance between them at time instant $t$ is less than their predefined headway distance. i.e.,
 \begin{equation}
  \label{eq:1.2.1:3}
   d_{jj'}^{\mathcal{H'}}<d_{jj'}^{\mathcal{H}}
  \end{equation} 
  
 \item Case 1.2.2: Train $T_j$ is standing at $p^{th}$ platform of station $S_i$ and train $T_{j'}$ is approaching to $S_i$ on the same platform at the same time $t$.
 \begin{equation}
  \label{eq:1.2.2:1}
  \begin{cases}
   P_{jpl}=1 \\
   L_{jl} ~and~ L_{j'l} = 1
  \end{cases}
 \end{equation}
 
 \begin{equation}
 \label{eq:1.2.2:2}
  \begin{cases}
   T_{j'}\rightarrow T_j  \\
  \vartheta_{j} = 0 \\
  \vartheta_{j'}>0
  \end{cases}
 \end{equation}
 
 
 \end{itemize}
 \end{itemize}
 With this background, a rear-end collision is detected if either equations (\ref{eq:1.2.1:1}) - (\ref{eq:1.2.1:3}) or
 equations (\ref{eq:1.2.2:1}) - (\ref{eq:1.2.2:2}) hold. For both the cases, it is assumed that the trains can communicate directly using their
 communication devices if $d_{jj'}^{H'}\leq2r$ and when one train $T_j$ is standing at station $S_i$, the other train
 $T_{j'}$ communicate directly with the station agent.
 
\subsection{Collision Resolution}
\label{resolution}
As described above, the problem is very challenging in real-time scenario. 
The conflict situation can be prevented in a distributed manner through message-passing among neighboring trains, stations, and junctions within the communication range.
In order to solve such problem, we represent each train, station, and junction as an autonomous agent. These agents are capable of communicating and coordinating
with their neighbors through message passing. We use the notion of \emph{max-sum} algorithm for decentralized coordination \cite{prob413, prob417} to solve such problem.
Here, agents negotiate by exchanging messages locally to achieve a desired solution globally.
Within the communication range all trains can communicate with each other and always take part in the collision detection-resolution task.
First the agents perform collision detection. In this phase all the agents check for a situation when the distance between two trains are less than their actual headway 
distance, which may lead to a fatal collision. If such scenario arises then collision resolution is needed to prevent the mishap.
Collision resolution is based on agent cooperation and negotiation. During collision detection phase, all the agents involved in collision scenario check 
for all the metric parameters: headway distance, braking distance, critical distance, and also their possible decision states. If more than one potential threats
are detected, then the most fatal collision is handled first and the lower one is handled later. The participating agents search for the safe state from
all possible set of states using the idea of max-sum algorithm as discussed \cite{prob413, prob417} in this section to detect collision. The proposed algorithm works
iteratively until a feasible solution is found. In each iteration, all the agents exchange their new 
modified state, generated by max-sum approach, with the neighboring agents through message passing. 
In this paper two possible state has been taken: \emph{move} further and \emph{stop} applying full brake.
If there are several alternatives for the agents, then the best possible solution is chosen depending on trains priority, minimum braking distance, and critical distance.
Finally all the agents acquire decided action of state and send the messages to the nearby train agents, station agents, and junction agents.
We first represent each agent as a function $\hat{U}$ (utility) and a variable $\nu$ (state), which are the vertices of the factor graph \cite{factor_graph}. The utility
of any agent $\hat{U}_z (\gamma_z) $ depends on its own state and the state of its neighbors, where, $\gamma=\{\nu_1 \ldots \nu_a\}$ and $z$ is
the total number of factors. So, the function node is connected with its own variable node and the variable nodes of its neighbors.

\begin{figure}[!htbp]
\centering
\subfigure[]{\label{fig:factor_graph1}\includegraphics[scale=0.35]{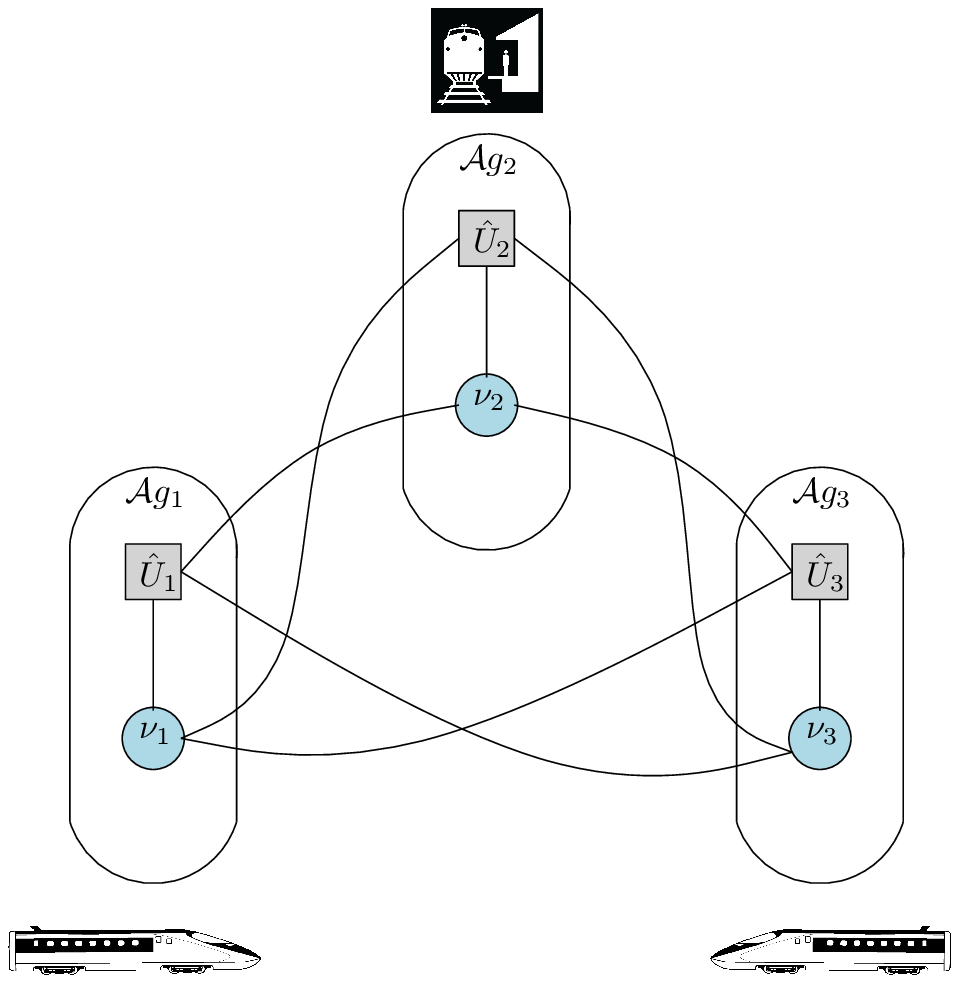}}
\hspace*{0.25cm}
\subfigure[]{\label{fig:factor_graph2}\includegraphics[scale=0.35]{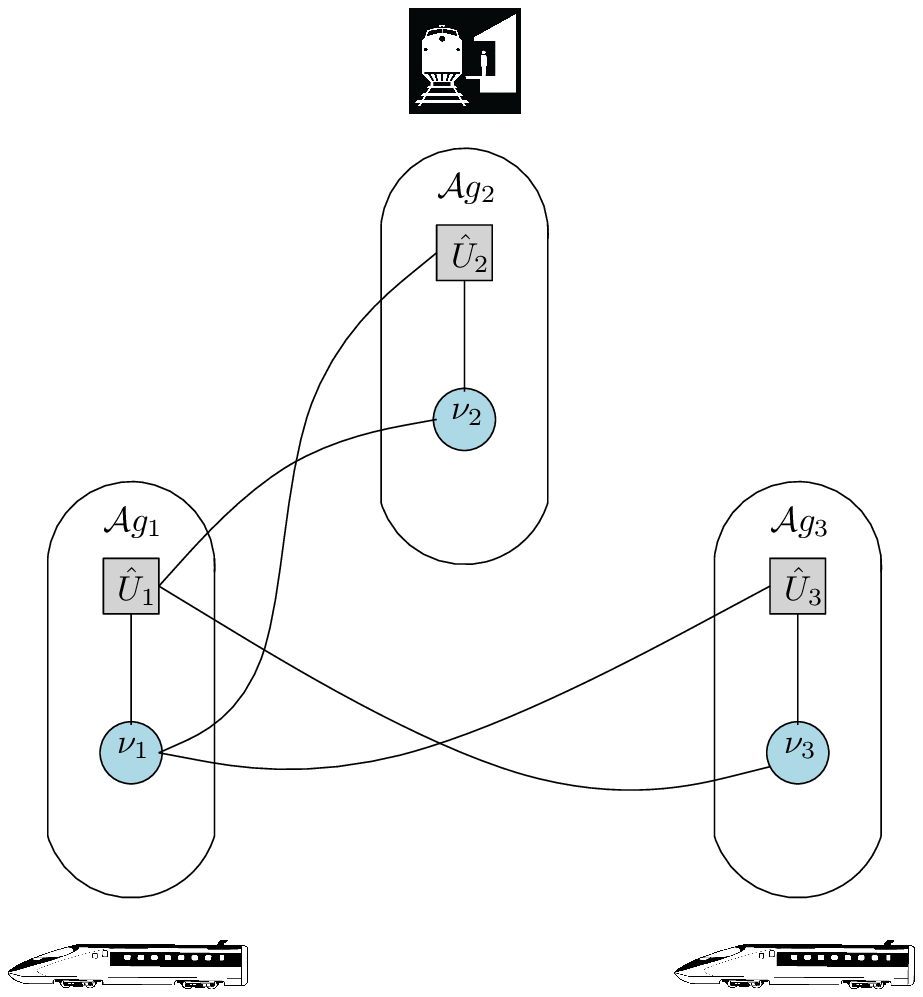}}
\captionsetup{justification=justified, singlelinecheck=false}
\caption{Factor graph representing agent communication. \newline
a for head-on collision.\newline
b for rear-end collision.}
\label{fig:factor_graph}
\end{figure}

Factor graphs in \emph{Figure \ref{fig:factor_graph}} describe the collision scenario with two train agents and one station agent or junction agent.
Here, $\mathcal{A}g_1$ and $\mathcal{A}g_3$ represent trains $T_1$ and $T_2$ participating in collision scenario. Whereas
$\mathcal{A}g_2$ represents the neighboring station 
$S_1$ or junction $\mathcal{J}_1$. Depending on the current distance between two trains $d_{jj'}^{\mathcal{H'}}$ the utility of individual $\mathcal{A}g_a$ varies.
In case of head-on collision (\emph{Figure \ref{fig:factor_graph1}}),
when the current distance between two trains $T_1$ and $T_2$ is greater than their communication range, then both train agents
$\mathcal{A}g_1$ and $\mathcal{A}g_3$ communicate with nearest station $\mathcal{A}g_2$. So, $\gamma_1 = \{\nu_1, \nu_2\}$, $\gamma_2 = \{\nu_1, \nu_2,\nu_3 \}$,
and $\gamma_3 = \{\nu_2, \nu_3\}$. Similarly, when two trains $T_1$ and $T_2$  are within their communication range, then train agents
$\mathcal{A}g_1$ and $\mathcal{A}g_3$ communicate through local message passing between themselves and nearest station agent $\mathcal{A}g_2$'s update 
depends on these two agents messages. So, $\gamma_1 = \{\nu_1, \nu_3\}$, $\gamma_2 = \{\nu_1, \nu_2,\nu_3 \}$, and $\gamma_3 = \{\nu_1, \nu_3\}$.
i.e.
when $d_{12}^{\mathcal{H'}}>2r$, $\sum_{z=1}^{3} \hat{U}_z (\gamma_z) = \hat{U}_1(\nu_1, \nu_2)+\hat{U}_2(\nu_1, \nu_2, \nu_3)+\hat{U}_3(\nu_2, \nu_3)$
and when $d_{12}^{\mathcal{H'}}\leq2r$, $\sum_{z=1}^{3} \hat{U}_z (\gamma_z) = \hat{U}_1(\nu_1, \nu_3)+\hat{U}_2(\nu_1, \nu_2, \nu_3)+\hat{U}_3(\nu_1, \nu_3)$.\\
Similarly, in case of rear-end collision (\emph{Figure \ref{fig:factor_graph2}}), as both the trains are within their communication range, i.e.
$d_{jj'}^{H'}\leq 2r$, $\mathcal{A}g_1$ and $\mathcal{A}g_3$ or $\mathcal{A}g_1$ and $\mathcal{A}g_2$ communicate directly with each other. So, either
$\gamma_1 = \{\nu_1, \nu_3\}$, $\gamma_2 = \{\phi \}$, $\gamma_3 = \{\nu_1, \nu_3\}$ and $\sum_{z=1}^{3} \hat{U}_z (\gamma_z) =
\hat{U}_1(\nu_1, \nu_3)+\hat{U}_3(\nu_1, \nu_3)$ or $\gamma_1 = \{\nu_1, \nu_2\}$, $\gamma_2 = \{\nu_1, \nu_2 \}$, $\gamma_3 = \{\phi\}$ 
and $\sum_{z=1}^{3} \hat{U}_z (\gamma_z) = \hat{U}_1(\nu_1, \nu_2)+\hat{U}_2(\nu_1, \nu_2)$.\\
We use max-sum in order to compute each agent's utility in distributed way, where,
\begin{equation}
 \mathring{U}_a (\nu_a)= \max_{\gamma - a} \sum_{a=1}^{z} \mathring{U}_a (\gamma_a)
\end{equation}
Again agent $a$'s optimal move $\nu_{a}^{*}$ is defined as,
\begin{equation}
 \nu_{a}^{*}=\arg \max_{\nu_a} \mathring{U}_a (\nu_a)
\end{equation}
Max-sum algorithm operates on two kind of functions,\\
$\bullet$ \underline{From variable to function:}\\
\begin{equation}
 \zeta_{\nu_a \rightarrow \hat{U}_z}(\nu_a) = \Phi_{az}+ \sum_{\substack{\hat{U}_{z'}\in neighbor(\nu_a) \\ \hat{U}_{z'}\neq\hat{U}_{z}}}
 \Gamma_{\hat{U}_{z'}\rightarrow \nu_a} (\nu_a)
\end{equation}
\noindent
$\bullet$ \underline{From function to variable:}\\
\begin{equation}
\Gamma_{\hat{U}_{z}\rightarrow \nu_a}(\nu_a) = \max_{\gamma - a} [\hat{U}_{z}(\gamma_z)+ \sum_{\substack{\nu_{a'}\in neighbor(\hat{U}_{z}) \\ \nu_{a'}\neq \nu_{a}}}
 \zeta_{\nu_{a'} \rightarrow \hat{U}_z} (\nu_{a'})]
\end{equation}
As described before, in distributed railway network, agents are located at each station ($SA$), train ($TA$), and junction point ($JA$).
Each agent selects an action for the train's state
from the set of possible actions \emph{move} and \emph{stop} to avoid collision.
Now, for each agent $\hat{U}_z(\gamma_z)$ is calculated as,
\begin{equation}
 \hat{U}_z(\gamma_z)=\beta_z(\nu_z) - \sum_{\substack{\nu_{z'}\in neighbor(\hat{U}_{z}) \\ \nu_{z'}\neq \nu_{z}}} (\nu_z \cdot \nu_{z'})
\end{equation}
where,
\begin{equation}
 \nu_z \cdot \nu_{z'}= 
 \begin{cases}
  1 ~~ if~ \nu_z=\mathcal{C}_R, ~\nu_{z'}=\mathcal{C}_R \\
  0 ~~otherwise
 \end{cases}
\end{equation}
$\beta_z(\nu_z)$ denotes the action state a train acquired at time instant $t$, that may lead to a collision.
For example w.r.t. \emph{Figure \ref{fig:factor_graph}}, let us consider that,

\begin{gather}
 \beta_1(\nu_1)=[-1, 1] \nonumber \\
 \beta_2(\nu_2)=[1, -1] \label{eq:beta_value}\\
 \beta_3(\nu_3)=[-1, 1] \nonumber
\end{gather}
where, equation (\ref{eq:beta_value}) shows current state of agents, taking part into a collision. 
Here the first column denotes \emph{stop} action and second column denotes \emph{move} action, where column
value $1$ means the train prefers to select that particular action at time instant $t$.
%

\section{Results and Discussion}
\label{results}
The simulation is coded using JAVA in UNIX platform of personal computer with 2.90 GHz processor speed and 4GB memory.
Proposed approach is compared against the existing centralized approach and with various other similar approaches 
\cite{prob411, wang2017novel, wu2015modeling, zhao2015positive} in JADE environment- a Java based agent development framework
\cite{jade}. The algorithms have been tested with varying number of trains, stations and junctions, taking real-time dataset from Indian Railways.
In this setup all the trains are currently standing either at stations or running on tracks throughout the railway network.
The speed of trains varies from 40 km/hr to 220 km/hr depending upon train category \cite{train_types_india}. Accordingly their braking distance
also vary. All trains maintain their speed during their journey. Headway distance of 200 m is taken to provide
collision free separation between trains at any time throughout the journey.
System efficiency is defined as a percentage proportion of number of collision detected and number of collision avoided by the 
proposed approach under specified framework and within 24 h time period.
\begin{equation*}
 system ~efficiency = \frac{no. ~of ~collision ~avoided}{no. ~of ~collision ~detected} \times 100\%
\end{equation*}

\begin{figure*}
\centering
\begin{minipage}{0.45\textwidth}
\centering
 \includegraphics[scale=0.5]{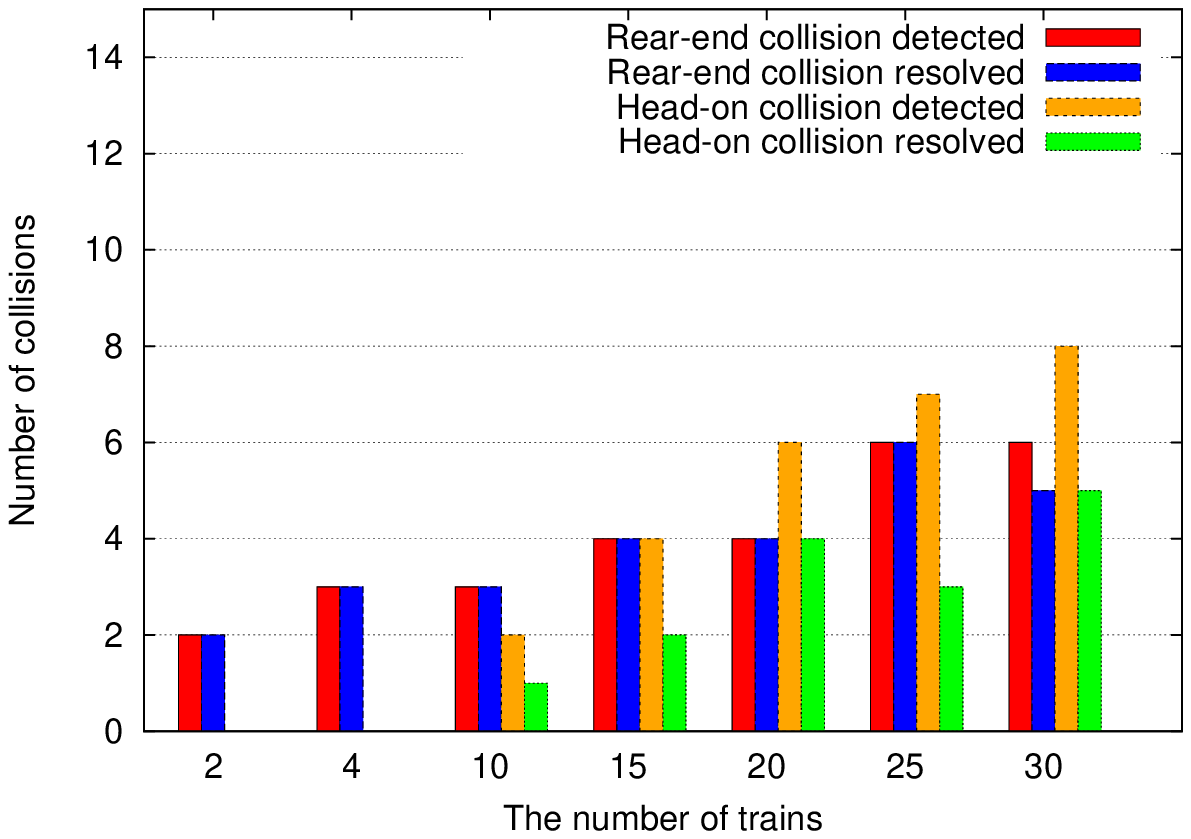}
 \captionsetup{justification=justified, singlelinecheck=false}
\caption{Number of collision detected and resolved using proposed approach for different number of trains.}
\label{fig:graph_detected_resolved}
\end{minipage}
\hspace{0.5cm}
\begin{minipage}{0.45\textwidth}
\centering
 \includegraphics[scale=0.5]{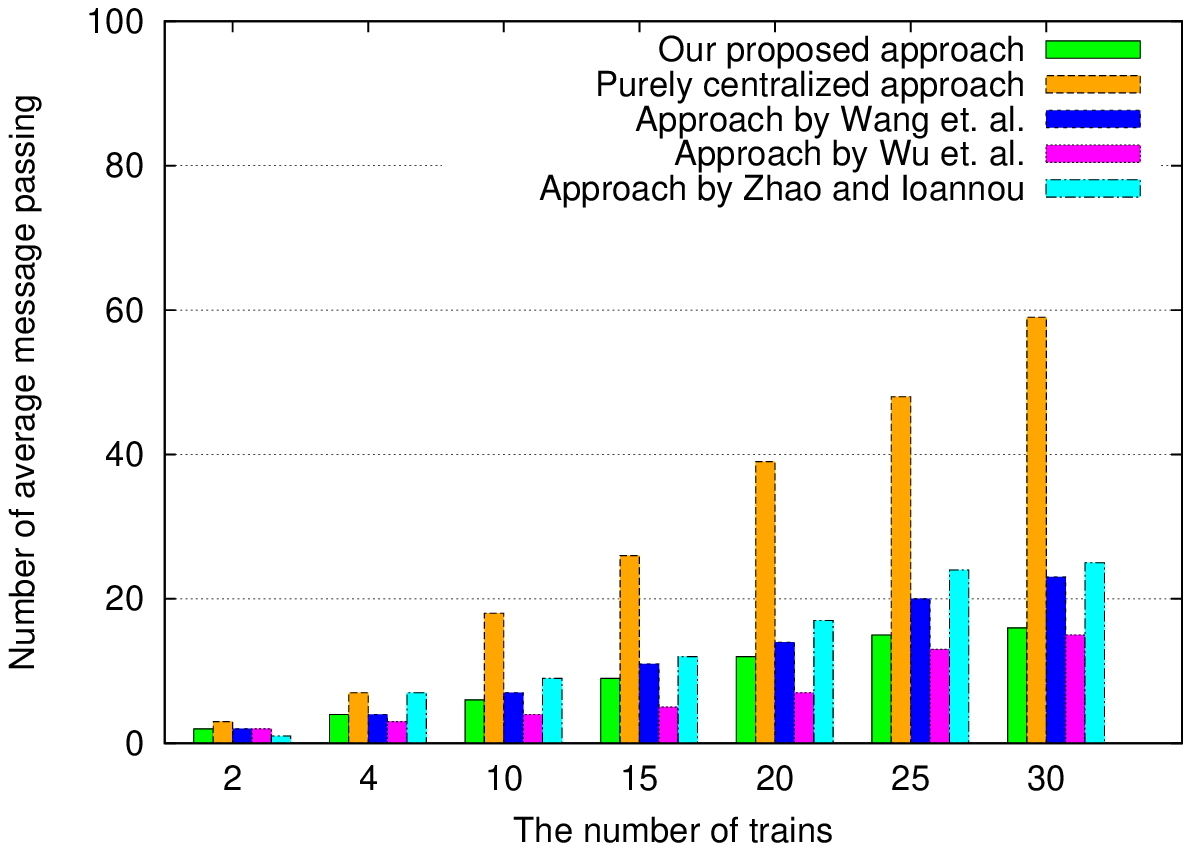}
 \captionsetup{justification=justified, singlelinecheck=false}
\caption{Average number of message passing among all the trains, stations, and junctions in given configuration to detect and resolve collisions.}
\label{fig:graph_msg_passing}
\end{minipage}
\begin{minipage}{0.45\textwidth}
\centering
\includegraphics[scale=0.5]{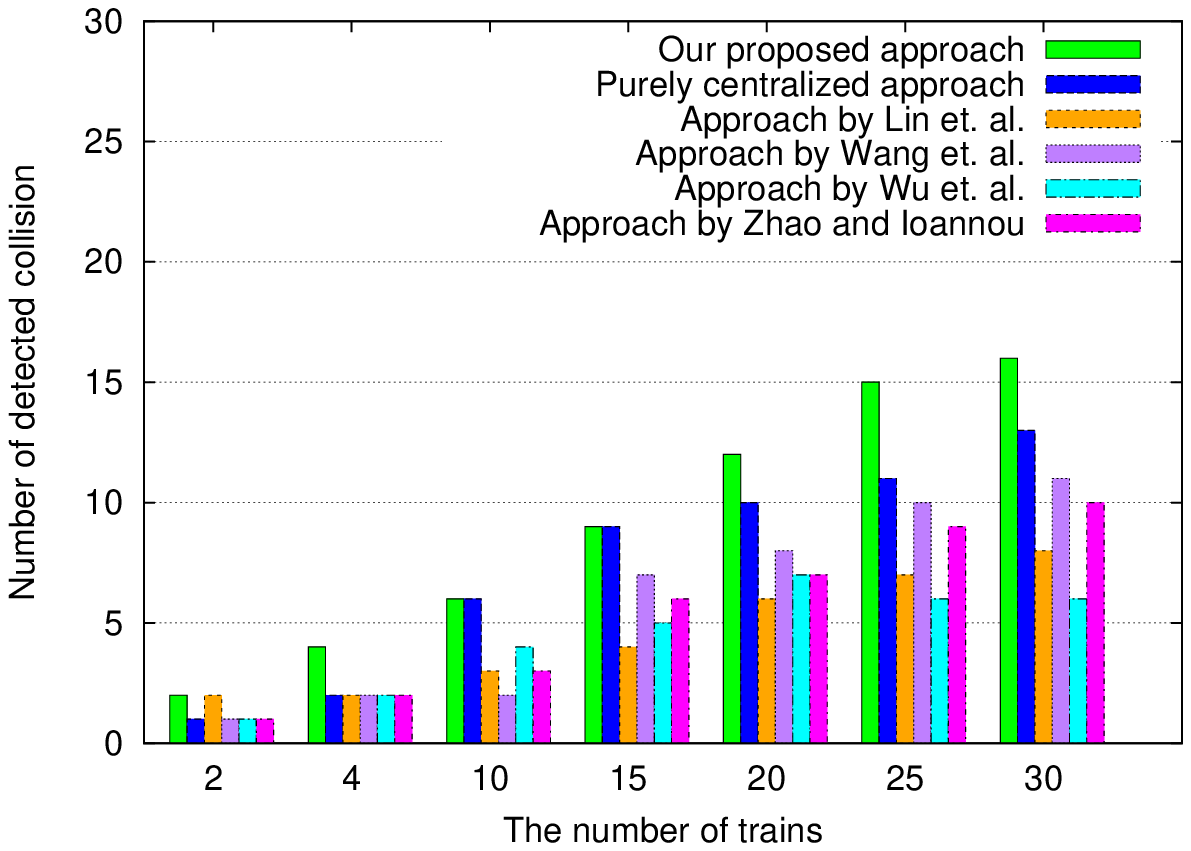}
 \captionsetup{justification=justified, singlelinecheck=false}
\caption{Number of collisions detected in proposed and various other approaches.}
\label{fig:graph_detection_actual_existing}
\end{minipage}
\hspace{0.5cm}
\begin{minipage}{0.45\textwidth}
\centering
\includegraphics[scale=0.5]{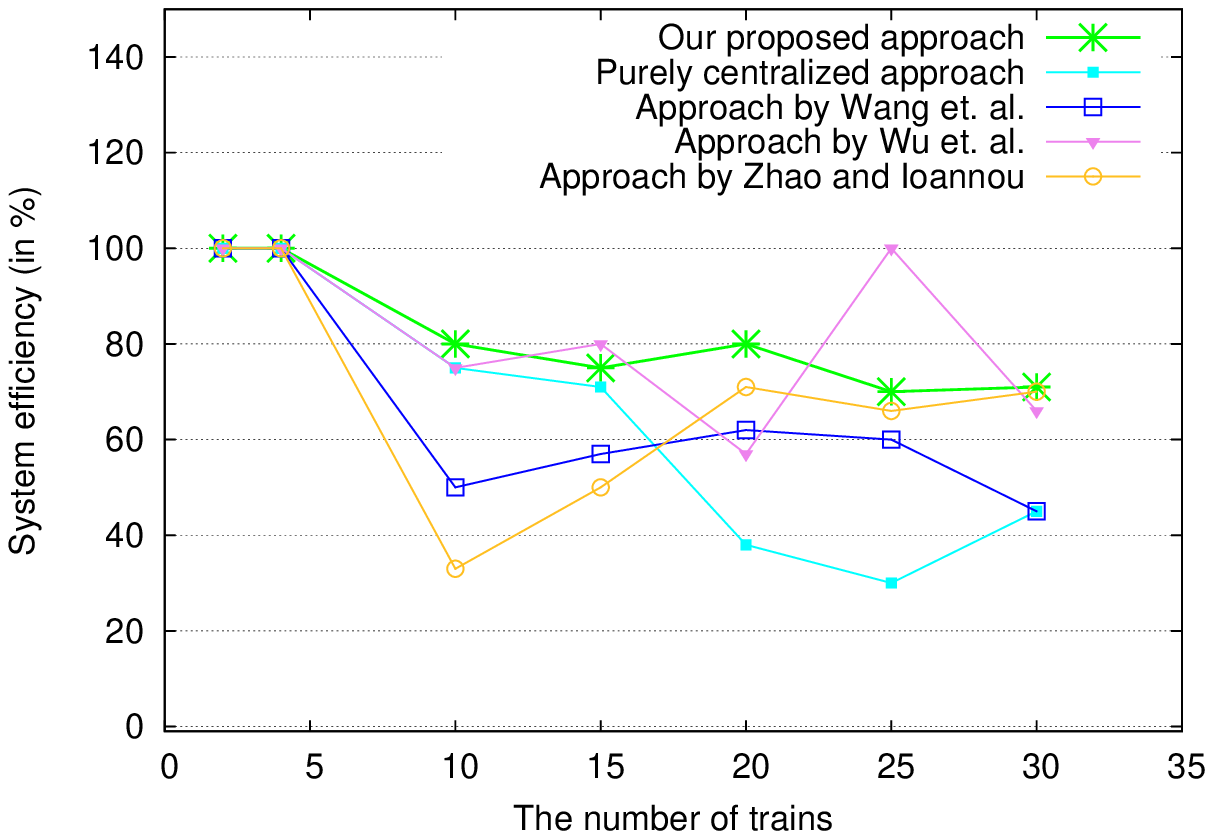}
 \captionsetup{justification=justified, singlelinecheck=false}
\caption{System efficiency achieved through the proposed approach and other existing approaches.}
\label{fig:graph_efficiency}
\end{minipage}
\end{figure*}

The first graph (\emph{Figure \ref{fig:graph_detected_resolved}}) shows the number of rear-end and head-on collision detected and resolved
by our proposed collision detection-resolution approach. The experiment is done for different number of trains (from 2 to 30)
for a particular time instant in the railway network, where number of stations and junctions are fixed for every setup.

The graph in \emph{Figure \ref{fig:graph_msg_passing}} presents the total number of message passing among all trains, stations, and junctions,
during collision handling procedure for proposed approach as well as other existing approaches cited in
paper \cite{wang2017novel, wu2015modeling, zhao2015positive}. In case of purely centralized approach there
is no train-to-train or junction-to-junction communication. Every train or junction is required to send information messages to nearby stations and the 
acknowledgement or decision messages from stations are sent to them accordingly to detect a collision each time it occurs. Hence the total number of information flow is 
comparatively high in this case. whereas, in case of proposed distributed approach, the trains, junctions, and stations communicate locally to detect a collision.
Hence a small number of messages are passed during this phase, shown in the above mentioned graph. Again, from this graph it is also noticed
that, compared to other existing techniques highlighted above, a very small number of messages are exchanged for the
collision handling technique in our proposed approach. 

\emph{Figure \ref{fig:graph_detection_actual_existing}} compares the collision detected by the
proposed detection approach and other existing approaches
discussed state-of-the-art \cite{prob411, wang2017novel, wu2015modeling, zhao2015positive} for the same experimental setup.
In these cited paper, depending on the communication strategies the total number of collision detection varies eventually.
It is clearly noticed that the number of detection are less than the actual
collision for the same setup. But in our proposed approach all types of agents,
i.e. train, junction, and station communicate with each other in a distributed manner to detect actual number of possible collisions.

\emph{Figure \ref{fig:graph_efficiency}} exhibits the system efficiency in various approaches
\cite{wang2017novel, wu2015modeling, zhao2015positive} in comparison with our proposed
approach for the same railway network setup. It is shown here that, for small number of trains (2 and 4) the system efficiency
is same for all the methods mentioned above. But when more trains are taken into consideration keeping other parameters static,
the efficiency varies in different methods. Using the approach, proposed in this paper, the efficiency is noticed to be much higher most 
of the cases. This validate our proposed work for using it in real-time railway scenarios.
 
\section{Conclusion}
\label{conclusion}

This paper addresses the problem of autonomous collision handling for trains in a large distributed complex railway system based on agent communication.
The proposed approach is divided into two parts: early detection of fatal collisions followed by collision resolution through avoidance. In this paper,
we showed how max-sum message passing algorithm can be applied to this domain to resolve potential collisions which are detected by system entities
(train, station, and junction). The presented concept considers all trains, stations, and junctions as autonomous agents, which can communicate and cooperate
during collision handling scenario. Proposed approach overcomes the need of repeated human interventions and control through track-side signaling, minimizing
chances of human error and/or signaling error. The basic idea of headway distance, braking distance, and critical distance have been taken to
determine the alarming situations which may lead to a collision.However, for the simplicity
of this approach it is assumed that all the railway entities are equipped with similar kind of communication devices having a certain range.
The internal mechanism of these devices are primarily out of scope of this paper.\\
\indent
We demonstrate the results using a railway network in Eastern Railway, India. Experimental evaluation shows that, the presented collision
handling approach provides better system efficiency as compared to other existing approaches. Number of average message passing is less with this approach,
which helps to minimize the overall communication cost. For the same railway network, our proposed approach can efficiently detect maximum number of 
collisions which might be overlooked by other existing approaches leading dreadful accidents.\\
\indent
The proposed approach can be deployed in all kinds of vehicles in railway domain. All the rail-wagons must be equipped with bi-directional
communication devices to provide all possible communication among them whenever needed. The solution mechanism can be integrated with the existing
railway safety management system and here lies the practicability of our presented approach.

\section*{Acknowledgement}
The authors are grateful for partial funding provided by DST and MHRD.
 
\bibliographystyle{unsrt}
\bibliography{prob4}   

\end{document}